\documentclass[reprint, secnumarabic,amssymb, nobibnotes, aps, prb]{revtex4-1}

\usepackage{color}
\usepackage{epsfig, graphics, graphicx, subfigure, wrapfig, lipsum}
\usepackage{verbatim}
\usepackage{multirow}
\usepackage{dcolumn}
\usepackage{bm}
\usepackage{amsmath, braket}
\usepackage{array}
\usepackage{xr}

\newcolumntype{X}[1]{>{\centering\arraybackslash\hspace{0pt}}p{#1}}
\newcolumntype{M}[1]{ >{\centering\arraybackslash}m{#1}}
\newcommand{\roml}[1]{\lowercase\expandafter{\romannumeral #1\relax}}
\newcommand{\romu}[1]{\uppercase\expandafter{\romannumeral #1\relax}}

\begin{document}

\title{Cross-Plane Thermal Transport in Layered Materials}

\author{Amey G. Gokhale}
\affiliation{Mechanical Engineering Department, IIT Bombay, India}
\author{Ankit Jain}
\affiliation{Mechanical Engineering Department, IIT Bombay, India}
\email{a_jain@iitb.ac.in}

\date{\today}

\begin{abstract}
The cross-plane (across-layers) phonon thermal transport of five diverse, layered semiconductors is investigated by accounting for higher-order four-phonon scattering,  phonon renormalization, and multi-channel thermal transport. For materials having relatively large cross-plane thermal conductivity ($\text{AlB}_6$, $\text{MoS}_2$, and $\text{MoSi}_2\text{N}_4$),  phonons contributing to cross-plane conductivity have an order of magnitude larger mean free path than that for the basal-plane thermal transport, whereas the opposite effect is observed for materials with low thermal conductivity ($\text{MoO}_3$  and $\text{KCuSe}$). 
The contribution from the wave-like coherent transport channel is less than 5\% in all considered materials. 
Our work unravels the contrasting role of nano-structuring on the basal- vs. cross-plane thermal conductivity of low and high thermal conductivity layered materials.
 
\end{abstract}

\maketitle

The tunable optoelectronic \cite{zhang2016van}, thermal \cite{gao2016lattice}, and electrical \cite{shukla2020tunable} properties of layered materials have made them a subject of continuous exploration for their potential applications in photonic devices\cite{liu20222d}, field-effect transistors\cite{roy2014field}, thermal management\cite{song2018two}, energy storage\cite{zhang20162d}, and  catalysis\cite{su20192d}. The layered materials are composed of different atomic layers that are held together by weak van der Waal (vdW) interactions \cite{levy2012intercalated}. While these interactions allow for easy manipulation of the structure, resulting in numerous homo/hetero-stacked structures \cite{geim2013van}, they also lead to inefficient thermal transport across layers, often causing performance bottlenecks in derived devices \cite{wei2011interfacial}. Extensive research focused on the understanding of basal-plane (parallel to atomic layers) thermal transport has resulted in  the discovery of new thermal transport physics, such as hydrodynamic thermal transport \cite{cepellotti2015phonon}, and ultrahigh/diverging thermal conductivity \cite{lindsay2013first,chen2020ultrahigh} in these materials.  The across-layer (cross-plane) thermal transport, however,  has received little attention despite its importance on the performance of emerging homo/hetero-stacked layered material-based devices \cite{hadland2019ultralow,zhou2022phonon}.

In this work, we investigate cross-plane thermal transport in five diverse, layered materials spanning hexagonal/orthorhombic, binary/ternary, intercalated/multi-layered structures using the state-of-the-art ab-initio calculations. We start by benchmarking vdW functionals for the study of thermal transport in layered materials and investigate phonon thermal transport by accounting for three- and four-phonon scatterings, phonon renormalization, and multi-channel thermal transport \cite{jain2020, jain2022}. We find that despite similar interlayer interactions, the cross-plane phonon thermal conductivity spans more than an order of magnitude, varying from  $0.48$ W/m-K for KCuSe to $11$ W/m-K for $\text{MoSi}_2\text{N}_4$ at room temperature. 
Further, even though cross-plane thermal transport is less efficient than the corresponding basal-plane transport in all considered materials, the phonons contributing to cross-plane transport can have longer mean free paths than those for basal-plane transport; thus suggesting a larger impact of nano-structuring on the cross-plane thermal compared to the basal-plane transport in some materials.

We obtain the phonon thermal conductivity, $k$, of considered semiconducitng materials  as \cite{mcgaughey2019,jain2020}:
\begin{equation}
 \label{eqn_k}
    k^{\alpha} = \sum_i c_{ph, i} v_{\alpha}^2 \tau_i^{\alpha},
\end{equation}
where the summation is over all the phonon modes in the Brillouin zone enumerated by $i\equiv(q,\nu)$, where $q$ and $\nu$ are phonon wavevector and mode index, and $c_{ph,i}$, $v_{\alpha}$, and $\tau_i^{\alpha}$ represent phonon specific heat, group velocity ($\alpha$-component), and transport lifetime respectively. The transport lifetimes are obtained by considering phonon-phonon scattering via three- and four-phonon scattering processes. The phonon-boundary scattering is included as $\tau_{bdry} = {L_{bdry}}/{|v|}$, where $L_{bdry}$ is the characteristic length scale (chosen as 1 $\mu$m for all materials).
The details regarding $c_{ph}$, $v_{\alpha}$, and $\tau$ calculations via phonon renormalization, temperature-dependent sampling of the potential energy surface, and multi-channel thermal transport are available elsewhere in Refs \cite{reissland1973,mcgaughey2019,jain2020}.

The computation of phonon mode properties (heat capacity, group velocity, three-phonon, and four-phonon scattering rates) requires harmonic, cubic, and quartic interatomic force constants \cite{jain2020} which we obtained using the density functional theory calculations as implemented in planewaves based quantum chemistry package Quantum Espresso \cite{giannozzi2009quantum,giannozzi2017advanced}. The structure relaxation and force constant calculations were carried out using Perdew–Burke-Ernzerhof Optimized Norm Conserving Vanderbilt  pseudo-potentials \cite{ernzerhof1999assessment, hamann2013optimized}. The plane wave energy cutoff and the electronic wavevector grid are converged to ensure a change in total energy to be less than $10^{-8}$ Ry/atom. The harmonic force constants are obtained using a finite-difference approach by displacing required atoms from their equilibrium positions by $0.05$ $\text{\AA}$.  The anharmonic force constants are obtained using the stochastic thermal snapshot technique as discussed in Ref.~\cite{Ravichandran2018} using 200 thermally populated supercells corresponding to a temperature of 300 K. The cubic interaction cutoff is fixed at  $4^{\text{th}}$ neighbor shell for the extraction of cubic force constants from the force-displacement data fitting. The obtained three-phonon thermal conductivities are converged within 5\% with respect to the above choice of interaction cutoffs and phonon wavevector grid for all considered materials. For the calculation of four-phonon scattering rates, the quartic interaction cutoffs are set at $1^{\text{st}}$ neighbor shell, and the phonon wavevector grid is selected to ensure a minimum number of 9200 phonon modes.

\begin{figure*}
\begin{center}
\epsfbox{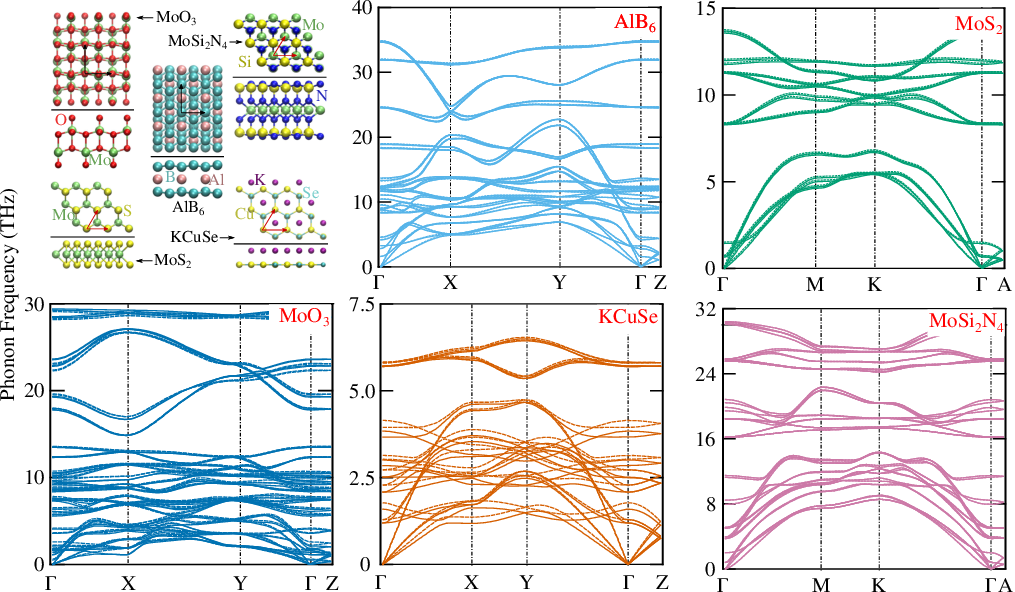}
\end{center}
\caption{{\bf Phonon Dispersions.} The phonon dispersions of considered layered materials were obtained by including quartic phonon renormalization at 300 K (dashed lines) and without renormalization (solid lines). The renormalization is significant for $\text{MoO}_3$ and KCuSe and is negligible in other materials. Compared to basal-plane directions, the optical phonon branches are relatively flat in the $\Gamma$-Z direction, suggesting narrow-range acoustic-phonon dominated phonon thermal transport in the cross-plane direction.}
\label{fig_dispersion}
\end{figure*}

 The crystal structures of considered layered materials are presented in Fig.~\ref{fig_dispersion}(a). 
$\text{AlB}_6$ consists of an atomic-layers/planes of boron atoms (borophene) which are lightly buckled and Al atoms are intercalated between these layers, resulting in a seven atoms unit-cell. 
The unit-cell of $\text{MoS}_2$ is composed of S and Mo atomic planes with each Mo atom bonded to two equivalent S atoms. In contrast, in $\text{MoO}_3$, (\roml{1}) each Mo atom is bonded to three different types of O atoms (singly, doubly, and triply coordinated) and (\roml{2}) $\text{MoO}_3$ belongs to orthorhombic crystal family compared to hexagonal for $\text{MoS}_2$. KCuSe has a similar structure as that of $\text{AlB}_6$, but guest atoms (K/Al in KCuSe/$\text{AlB}_6$) are sandwiched in-between planar binary CuSe layers as opposed to non-planar monoatomic layers in $\text{AlB}_6$. $\text{MoSi}_2\text{N}_4$ has a heterogeneous stacking with $\text{MoN}_2$ layers sandwiched between two $\text{SiN}$ layers resulting in three atomic layers thick primitive unitcell.

\begin{table*}
\begin{center}
\caption{{\bf vdW Functional Benchmarking.} The effect of selected vdW functionals on the predicted structural, harmonic, and anharmonic properties of monolayers of considered layered materials. The reported lattice constants ($a$) are for basal-plane, and the reported group velocity ($\overline{|v|}$) and Gr\"{u}neissen parameter ($\overline{\gamma}$) are heat capacity weighted averages of all phonon modes.  }
\label{table_vdW}
\begin{tabular}{c|c|c|c|c|c|c|c}
    \hline
    \multicolumn{2}{c|}{} & \multicolumn{4}{c|}{monolayer} & \multicolumn{2}{c}{bulk}\\
    \hline
    Material  & vdW  & \multicolumn{2}{c|}{basal-plane}  & $\overline{|v|}$   &  $\overline{\gamma}$   &   \multicolumn{2}{c}{cross-plane} \\
     & functional & \multicolumn{2}{c|}{lattice constant (\AA)} & (m/s) & & \multicolumn{2}{c}{lattice constant (\AA)}\\
    \hline
    & & exp. & pred. & & & exp. & pred.  \\
    \hline

    \multirow{3}{*}{$\text{AlB}_6$} & No & \multicolumn{1}{c|}{\multirow{3}{*}{-}} & 2.92 (3.42) & 3714 & 0.88 & \multicolumn{1}{c|}{\multirow{3}{*}{-}} & 14.02    \\
                              & D3 &        & 2.92 (3.42) & 3726      & 0.92 & & 13.23      \\
                              & rvv10  &    & 2.93 (3.43) & 3655      & 0.92  & & 13.28   \\
    \hline
    \multirow{3}{*}{$\text{MoS}_2$}  & No & \multicolumn{1}{c|}{\multirow{3}{*}{3.16\cite{wieting1971infrared}}}  & 3.19   & 1818    & 1.29 &        &  14.07 \\
                              & D3     &    & 3.17   & 1831    & 1.30  & 12.29\cite{wieting1971infrared} & 12.41  \\
                              & rvv10  &    & 3.22   & 1787    & 1.32 &       & 12.36  \\
    \hline
    \multirow{2}{*}{$\text{MoO}_3$}     & No  & \multicolumn{1}{c|}{\multirow{2}{*}{3.70 (3.92)\cite{sian2004optical}}}      & 3.68 (3.93)  & 1990      & 1.86   &  & - \\
                              & rvv10    &  & 3.70 (3.92) & 1956      & 1.82 & 13.35\cite{sian2004optical} & 14.00\\
    \hline
    \multirow{3}{*}{KCuSe}    & No  &  \multicolumn{1}{c|}{\multirow{3}{*}{4.18\cite{savelsberg1978ternare}}}     & 4.16        & 848       & 1.53   & & 9.95  \\
                              & D3   &      & 4.13        & 876       & 1.61 & 9.54\cite{savelsberg1978ternare} & 9.98     \\
                              & rvv10  &    & 4.18        & 823       & 1.75  &  & 9.59  \\
    \hline
    \multirow{3}{*}{$\text{MoSi}_2\text{N}_4$}  & No  &       & 2.91         & 3026      & 0.87 & \multicolumn{1}{c|}{\multirow{3}{*}{-}}  & 20.77   \\
                              & D3   &  2.94\cite{hong2020chemical}    & 2.90        & 3038      & 0.82  &  & 20.15  \\
                              & rvv10  &    & 2.92        & 3048      & 0.85  &  & 20.10  \\
    \hline

\end{tabular}
\end{center}
\end{table*}

We start by first validating the choice of employed vdW functional to describe weak interlayer interactions in layered materials. For this, we employed a similar methodology as we used in our previous work for studying thermal transport in $\text{MoS}_2$ \cite{gokhale2021cross}.  In brief, we started with several empirical post-DFT corrections and non-local vdW functionals and benchmarked them for (\roml{1}) basal-plane lattice constant prediction as compared to that without employing any vdW correction and (\roml{2}) phonon group velocities and anharmonicty (as measured by heat capacity weighted Gr\"{u}neissen parameters) against those predicted without any vdW correction for monolayer material.
Amongst all considered corrections/functionals, we found that only Grimme-D3 \cite{grimme2010consistent} post-DFT correction and rvv10 \cite{vydrov2010nonlocal} non-local functional were able to correctly describe the considered properties of $\text{MoS}_2$, and as such, only these functionals are considered in this study. 

For mono-layer materials, the computationally predicted basal-plane lattice constant, the phonon heat-capacity weighted group velocity, and Gr\"{u}neissen parameter are reported  in Table ~\ref{table_vdW}. For monolayer $\text{MoO}_3$, Grimme-D3 correction-based DFT calculations did not converge, and these results are not available in Table ~\ref{table_vdW}. For all considered materials, the predicted basal-plane lattice constant and heat-capacity weighted group velocities using the Grimme-D3 and rvv10 functional are within  {2}\% of those obtained using the no-vdW functional. Furthermore, with the exception of the rvv10 functional for KCuSe, 
both functionals  produced less than a 5\% difference in Gr\"{u}neissen parameters compared to those obtained without vdW functional for all materials. Therefore, other than $\text{MoO}_3$, Grimme-D3 functional is employed to study thermal transport across layers in all materials in this study. For $\text{MoO}_3$, since monolayer calculations did not converge with Grimme-D3,  the rvv10 functional is employed. We find that with an exception of inter-layer spacing in $\text{MoO}_3$, our predicted lattice constants (both basal-plane and cross-plane directions) are within 2\% of experimentally measured values \cite{sian2004optical} for these choices of functionals.

The phonon dispersions of considered materials are reported in Fig.~\ref{fig_dispersion}(b)-\ref{fig_dispersion}(f). 
Due to intercalation of Al, the phonon frequency range in $\text{AlB}_6$ is reduced compared to that reported in literature for ${\delta}_4$-borophene \cite{hu2020three}.
For $\text{MoS}_2$ and $\text{MoO}_3$, the acoustic phonon modes are dominated by heavy Mo atoms with lower phonon frequencies compared to that for Al-dominated modes in  $\text{AlB}_6$.
In KCuSe, in addition to the presence of heavy and intercalated atoms, the interatomic planar bonding is also weak. This weak bonding is reflected in thermal mean square displacements (MSD), which are high at  $0.03$ and $0.04$ $\text{\AA}^2$ for K and Se atoms in the basal- and cross-plane directions respectively at 300 K compared to less than $0.02$ $\text{\AA}^2$ for all other materials. As such, the sound speed and the phonon frequency range are less than $2500$ m/s  and 7 THz in KCuSe. For $\text{MoSi}_2\text{N}_4$, despite heavy Mo atoms, the sound speed is larger than 6000 m/s in the $\Gamma$-Z direction owing to strong interatomic bonds formed by N atoms. 

Because of large thermal displacements of atoms in KCuSe, the potential energy surface experienced by atoms at non-zero temperatures is different from that at 0 K, which results in a strong phonon renormalization. The phonon dispersions of considered materials obtained by phonon renormalization at 300 K are reported as dashed lines in Fig.~\ref{fig_dispersion}. As expected from large MSDs of atoms, the renormalization is strong in KCuSe (also for $\text{MoO}_3$ with MSD of up to $0.02$ $\text{\AA}^2$) and is negligible in other materials (MSD are less than $0.01$ $\text{\AA}^2$ in all other materials) \cite{MoSi2N4_note}. 


\begin{figure*}
\begin{center}
\epsfbox{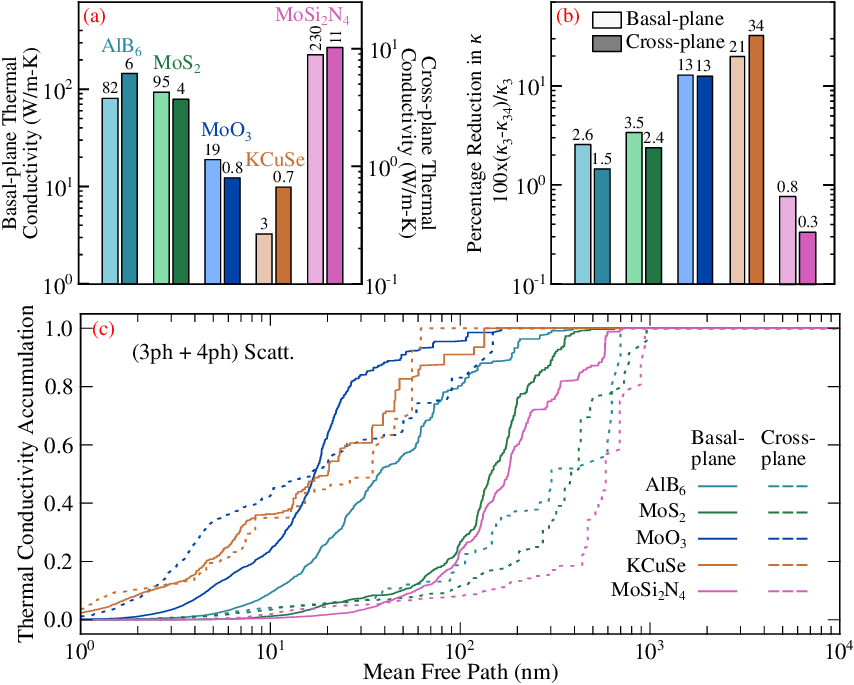}
\end{center}
\caption{{\bf Phonon Thermal Conductivity.} 
(a) The predicted thermal conductivities of considered materials obtained by considering three-phonon scattering, (b) percentage reduction in predicted thermal conductivities on inclusion of four-phonon scattering, and (c) thermal conductivity accumulation with phonon mean free paths (by considering both three- and four-phonon scatterings). Only zigzag-direction values are reported for the basal-plane. For KCuSe and $\text{MoO}_3$, the results are obtained by considering phonon quartic renormalization. All results are obtained by including phonon-boundary scattering corresponding to the characteristic sample size of 1 $\mu$m. Other than KCuSe and $\text{MoO}_3$, the thermal transport in the cross-plane direction is dominated by longer mean free path phonons. }
\label{fig_thermalK}
\end{figure*}

The predicted basal-plane (zigzag direction) and cross-plane thermal conductivities of considered materials obtained by considering three-phonon scattering only  are reported in Fig.~\ref{fig_thermalK} (a). The reported thermal conductivities are obtained using the relaxation time approximation 
(RTA) of the BTE, and the effect of the iterative solution of the BTE on the cross-plane (basal-plane) thermal conductivity is found to be less than 5\% (17\%) for all considered materials at 300 K using only three-phonon scattering. The obtained basal- and cross-plane thermal conductivities of considered materials span over an order of magnitude and follow a similar relative trend between considered materials. The lowest obtained thermal conductivities are for KCuSe due to weakly bonded atoms (and hence low group velocities) and strongly anharmonic bonds (strong phonon-phonon scattering) as reported in Table~\ref{table_vdW}. Amongst $\text{MoS}_2$ and $\text{MoO}_3$, both having similar group velocities, the thermal conductivities are higher for $\text{MoS}_2$ due to higher structural symmetry (all Mo-S bonds are similar in $\text{MoS}_2$ compared to three different kind of Mo-O bonds in $\text{MoO}_3$) and relatively weak anharmonicity.
For $\text{AlB}_6$, while the group velocities are larger and the anharmonicity is lower than that in $\text{MoS}_2$, the thermal conductivity is lower due to less symmetric orthorhombic structure with enhanced phonon scattering phase space. 
For $\text{MoSi}_2\text{N}_4$, the thermal conductivities are high at 230 and 11 W/m-K in the basal- and cross-plane directions due to large phonon group velocities, weak anharmonicity, and symmetric structure.

The inclusion of four-phonon scattering reduces phonon lifetimes, thus reducing thermal conductivity.  Fig.~\ref{fig_thermalK} (b) presents the percentage reduction in predicted thermal conductivities when four-phonon scattering is considered. For KCuSe and $\text{MoO}_3$, the reported results are obtained by including phonon renormalization. Considering only three-phonon scattering, the cross-plane thermal conductivity is under-predicted by {42}\% 
at 300 K for KCuSe if phonon renormalization is not included.

For basal-plane thermal transport, the four-phonon scattering has a small effect on the thermal conductivity, and the maximum obtained reduction in thermal conductivity is 21\% 
for KCuSe with strongly anharmonic bonds. For cross-plane thermal transport, while the effect is more pronounced and the thermal conductivity is reduced by 13\%  
for $\text{MoO}_3$ and 34\% 
for KCuSe, the effect is significant only for strongly anharmonic materials ($\text{MoO}_3$ and KCuSe). For $\text{AlB}_6$, $\text{MoS}_2$, and $\text{MoSi}_2\text{N}_4$ with strongly bonded atoms, the thermal conductivity remains unchanged (change is less than 5\%) with the inclusion of four-phonon scattering.

For $\text{MoS}_2$, our predicted basal-plane thermal conductivity obtained by iteratively solving the Boltzmann transport equation is 111  W/m-K {(95 using the RTA solution of the BTE)} for 1 $\mu$m sample, directly comparable with the experimentally measured values of 85-110 W/m-K reported by Liu et al. \cite{liu2014measurement}. Likewise, our predicted cross-plane thermal conductivity of $\text{MoS}_2$ is 4 W/m-K, is confined in the experimentally measured thermal conductivity range 4.4 $\pm$ 0.45 W/m-K by Jiang et al.~\cite{jiang2017probing}. For $\text{MoO}_3$, our predicted average basal-plane thermal conductivity obtained by considering only three phonon scattering for a sample size of 1 $\mu$m is 17.7 W/m-K compared to $13.9$  W/m-K by Tong et al.~\cite{tong2021ultralow} using similar calculations. {Amongst other numerical parameters, this difference can be due to the non-inclusion of phonon renormalization in the study of Tong et al; by considering only three-phonon scattering, we find that the average basal-plane thermal conductivity of $\text{MoO}_3$ is under-predicted by 7\% without phonon renormalization.

While all considered materials have similar thermal conductivity trends,  we found that the phonon modes that contribute to basal- and cross-plane transport have different behavior in low- and high-thermal materials [Fig.~\ref{fig_thermalK}(c)]. 
For materials having low cross-plane thermal conductivity (KCuSe and $\text{MoO}_3$), the  phonons contributing to basal- and cross-plane thermal transport have similar mean free path ranges. However, for remaining materials with high cross-plane thermal conductivity, the phonon mean free paths for the cross-plane thermal transport are larger than those for the basal-plane direction. This is understandable from phonon dispersion where noticeably, other than low-frequency acoustic phonon branches, all other branches are dispersionless in the $\Gamma$-Z direction in $\text{AlB}_6$, $\text{MoS}_2$, and $\text{MoSi}_2\text{N}_4$. Thus, all contribution to cross-plane thermal transport comes from acoustic phonons  with narrow frequency  range, which consequently results in a narrow phonon mean free path range for the cross-plane direction compared to that for the basal-plane direction. Furthermore, since acoustic phonons have, in general, larger group velocities compared to the optical phonons \cite{dove1993introduction}, the phonon mean free paths for the cross-plane thermal transport are larger than those for the basal-plane direction in these materials. For KCuSe and $\text{MoO}_3$, the weak interatomic bonding and strong anharmonicity result in small phonon mean free paths for acoustic phonons, and the  contribution is similar from acoustic and optical phonons in the basal- and cross-plane directions. As a result, the mean free paths of heat-carrying phonons in the cross-plane direction are comparable with the basal-plane direction for these low  thermal conductivity materials.

\begin{figure}
\begin{center}
\epsfbox{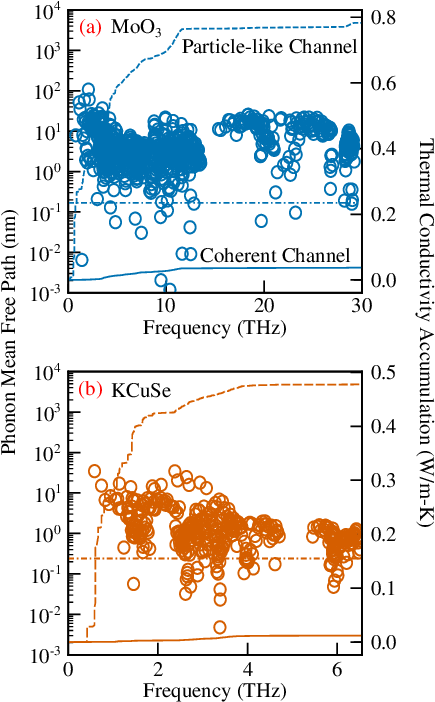}
\end{center}
\caption{{\bf Multi-channel Thermal Transport.} The comparison of phonon mean free paths for the cross-plane thermal transport with minimum inter-atomic spacing (Ioffe-Regel limit \cite{Ioffe-Regel}) for (a) $\text{MoO}_3$ and (b) KCuSe. The contribution of different frequency modes towards particle-like phonon and wave-like coherent transport channels are plotted on the secondary axes. The contribution from the coherent channel is less than {5\%}  for the cross-plane thermal transport in all considered materials.}
\label{fig_coherent}
\end{figure}

The obtained cross-plane thermal conductivities of $\text{MoO}_3$ and KCuSe are lower than {1} $\text{W/m-K}$ at room temperature. For such low thermal conductivity materials, other than the contribution from a particle-like transport channel, an additional contribution originating from the wave-like coherent channel can also be significant for thermal transport \cite{jain2022single}. To check for this, the phonon mean paths are compared with minimum interatomic distance (Ioffe-Regel limit\cite{Ioffe-Regel}) in Fig.~\ref{fig_coherent}. We find that for all considered materials, the cross-plane thermal transport is dominated by phonons with mean free paths larger than the Ioffe-Regel limit. Therefore, despite the low thermal conductivity from particle-like transport channels, the contribution from coherent-channel is expected to be small \cite{jain2022single}. This is indeed the case, and the contribution of the coherent channel is obtained to be less than {5}\% for all considered materials in the cross-plane direction. 
For the basal-plane thermal transport, the contribution from the coherent channel is even less (less than {2}\% for all considered materials). Therefore, though the basal- and cross-plane thermal conductivities of considered 2D materials span over several orders of magnitude, the thermal transport is only through the particle-like single transport channel.

In summary, we investigated cross-plane thermal transport in five diverse, layered materials using ab-initio-driven density functional theory calculations. 
We find that four-phonon scattering is only needed to correctly describe the cross-plane thermal transport in strongly anharmonic layered materials, even though cross-plane interactions are weak in all considered materials. The cross-plane thermal conductivity spans more than an order of magnitude, and the thermal transport is only through the particle-like single channel. Further, we find that the thermal transport in the cross-plane direction is majorly dominated by acoustic phonons and is more sensitive to material thickness owing to longer mean free paths of contributing phonons in high thermal conductivity materials, 
{while for low thermal conductivity materials, the mean free paths of heat-carrying phonons in the basal-plane direction are comparable with phonons in the cross-plane direction.} Our study highlights the different roles of nanostructuring on the cross- and basal-plane thermal transport in low- vs. high-thermal conductivity layered materials.

The authors acknowledge the financial support from the National Supercomputing Mission, Government of India (Grant Number: DST/NSM/R\&D-HPC-Applications/2021/10) and Core Research Grant, Science \& Engineering Research Board, India (Grant Number: CRG/2021/000010).  The calculations are carried out on the SpaceTime-II supercomputing facility of IIT Bombay and the PARAM Sanganak supercomputing facility of IIT Kanpur.

The raw/processed data required to reproduce these findings are available on a reasonable request via email.


\begin{thebibliography}{40}
\bibitem [{MoS()}]{MoSi2N4_note}%
  \BibitemOpen
  \href@noop {} {}\bibinfo {note} {Our phonon renormalization calculation did
  not converge for $\text{MoSi}_2\text{N}_4$. The effect is, however, expected
  to be minimal in $\text{MoSi}_2\text{N}_4$ due to small MSDs.}\BibitemShut
  {Stop}%
\bibitem [{Iof()}]{Ioffe-Regel}%
  \BibitemOpen
  \href@noop {} {}\bibinfo {note} {The Ioffe-Regel limit is obtained as the
  minimum interatomic spacing in the lattice.}\BibitemShut {Stop}%
\end{thebibliography}

%

\end{document}